%% file: main.tex
\documentclass[sigconf]{acmart}

\usepackage{soul}
\usepackage{enumitem}
\usepackage{makecell}

\copyrightyear{2025}
\acmYear{2025}
\setcopyright{cc}
\setcctype{by}
\acmConference[CHI '25]{CHI Conference on Human Factors in Computing Systems}{April 26-May 1, 2025}{Yokohama, Japan}
\acmBooktitle{CHI Conference on Human Factors in Computing Systems (CHI '25), April 26-May 1, 2025, Yokohama, Japan}
\acmDOI{10.1145/3706598.3714047}
\acmISBN{979-8-4007-1394-1/25/04}


\begin{document}

\pdfstringdefDisableCommands{\def\\{ }}
\title[An Empirical Study of Updating Survey Articles in Computing Research]{Toward Living Narrative Reviews:\\ An Empirical Study of the Processes and Challenges in Updating Survey Articles in Computing Research}

\author{Raymond Fok}
\affiliation{
  \institution{University of Washington}
  \city{Seattle}
  \state{WA}
  \country{USA}
}
\email{rayfok@cs.washington.edu}

\author{Alexa Siu}
\affiliation{
  \institution{Adobe Research}
  \city{San Jose}
  \state{CA}
  \country{USA}
}
\email{asiu@adobe.com}

\author{Daniel S. Weld}
\affiliation{
  \institution{Allen Institute for AI \&}
   \institution{University of Washington}
  \city{Seattle}
  \state{WA}
  \country{USA}
}
\email{danw@allenai.org}


\begin{abstract}
    \input{sections/00_abstract}
\end{abstract}



\maketitle

\input{sections/01_introduction}
\input{sections/02_related_work}
\input{sections/03_methods}
\input{sections/04_findings}
\input{sections/06_discussion}
\input{sections/07_conclusion}

\bibliographystyle{ACM-Reference-Format}
\bibliography{main}

\appendix
\input{sections/08_appendix}

\end{document}

%% file: sections/00_abstract.tex
Surveying prior literature to establish a foundation for new knowledge is essential for scholarly progress. However, survey articles are resource-intensive and challenging to create, and can quickly become outdated as new research is published, risking information staleness and inaccuracy. Keeping survey articles current with the latest evidence is therefore desirable, though there is a limited understanding of why, when, and how these surveys should be updated. Toward this end, through a series of in-depth retrospective interviews with 11 researchers, we present an empirical examination of the work practices in authoring and updating survey articles in computing research. We find that while computing researchers acknowledge the value in maintaining an updated survey, continuous updating remains unmanageable and misaligned with academic incentives. Our findings suggest key leverage points within current workflows that present opportunities for enabling technologies to facilitate more efficient and effective updates.

%% file: sections/01_introduction.tex
\section{Introduction}
Reviewing and building upon existing knowledge is fundamental to scholarly research. However, with the acceleration of research production and publication, keeping at the forefront of all available literature has become increasingly complex. Literature reviews offer tremendous value in synthesizing research, but are time-consuming and resource-intensive to create~\cite{tricco2008following, michelson2019significant}. Moreover, they quickly become outdated as new research emerges~\cite{Shekelle2001ValidityOT}. One survival analysis suggests some reviews may be outdated by the time of their publication and over 25\% require updating just two years after publication~\cite{shojania2007quickly}. Such information staleness presents a threat to validity, potentially under-informing researchers about research opportunities or critically misleading decision-makers.

In response, researchers have explored the concept of ``living reviews''---documents that are continually updated as new evidence emerges~\cite{elliott2014living, wijkstra2021livinglitreviews, cochrane2019handbook}, typically centered around living \textit{systematic} reviews. Maintaining living reviews can be challenging, however, with studies finding many are never updated after initial publication~\cite{tricco2008following, heron2023update}. Prior work on living reviews has primarily focused on accelerating literature discovery and appraisal~\cite{thomas2017living, vergara2020living}, but recent advances in artificial intelligence (AI) and large language models suggest an opportunity for supporting more complex cognitive aspects of updating reviews, such as evidence synthesis~\cite{martinboyle2024shallow, wang2024autosurvey}. Such capabilities are particularly relevant for living \textit{narrative} reviews, such as surveys in computing research, which rely more on expert interpretation and conceptual synthesis than the standardized protocols and meta-analyses of systematic reviews~\cite{snyder2019literature}.

To examine how narrative reviews are created and maintained in practice, we conducted semi-structured, retrospective interviews with 11 survey authors across diverse areas of computing research. Authors reflected on their processes and points of friction throughout their authoring and revision workflows, and shared their perspectives on when, how, and with what content these reviews should be updated. In understanding these practices, we sought to identify opportunities to support the updating process, including the potential role of AI assistance.

Our findings reveal the varied methodologies used and challenges encountered when authoring and updating survey articles, especially in paper discovery, taxonomy development, and synthesis. We identify three key types of updates for maintaining narrative reviews: 1) empirical updates involving evidence and examples, 2) structural updates to taxonomies and paper organization, and 3) interpretive updates to syntheses and framing. Each type offers distinct opportunities for AI assistance, from routine tasks like recalculating numerical values to more complex support for identifying potential biases and emerging research gaps. While authors saw the potential of AI to assist with routine tasks, they were skeptical of its ability to handle more nuanced interpretive tasks like constructing a compelling narrative. Moreover, the subjective, expertise-driven nature of their workflows raises important considerations for future tools enabling living narrative reviews.

%% file: sections/02_related_work.tex
\section{Background and Related Work}
There exist different types of literature reviews, varying across their goals, processes, and degrees of systematicity, with one typology identifying 14 distinct review types (e.g., critical, scoping, systematic, qualitative synthesis, and umbrella reviews)~\cite{grant2009typology}. More broadly, reviews can be categorized into one of three major approaches: systematic, semi-systematic, or integrative~\cite{snyder2019literature}. Systematic literature reviews, developed for and commonly utilized in clinical medicine, use ``explicit, systematic methods to identify, appraise, and synthesize all the empirical evidence that meets pre-specified eligibility criteria to answer a specific research question.''\footnote{\url{https://www.cochranelibrary.com/about/about-cochrane-reviews}} As the gold standard of evidence synthesis for clinical decision-making, systematic review methods have been studied extensively (e.g.,~\cite{chandler2013cochrane, pati2018write, lasserson2019starting}). On the other hand, semi-systematic or \textit{narrative} reviews are preferable when reviewing every relevant article is not possible. These reviews may rely on quasi-systematic search and appraisal methods, combine both qualitative and qualitative evidence, and hold broader aims beyond a specific research question, such as to survey a research area or track its progression over time. Closely related to narrative reviews, integrative or critical reviews further critique the literature to develop new theoretical perspectives, often integrating heterogeneous sources beyond peer-reviewed research articles to provide a synthesis that goes beyond descriptive or historical.

Creating these different reviews to synthesize the literature is often time-consuming and costly~\cite{tricco2008following, michelson2019significant}. Researchers have therefore explored the potential for AI to support literature review creation, among other research activities~\cite{van2023ai, morris_scientists_2023, messeri2024artificial}. For instance, prior work has developed approaches to automatically scaffold (e.g.,~\cite{hsu-etal-2024-chime, newman-etal-2024-arxivdigestables}) and generate review sections (e.g.,~\cite{li-ouyang-2024-related, bolanos2024artificial, wang2024autosurvey}), and proposed interactive methods to support relevant processes such as literature discovery (e.g.,~\cite{Kinney2023TheSS, xiao2023AutoSurveyGPT}) and appraisal (e.g.,~\cite{kusa2023cruise, chai2021research}), comprehension (e.g.,~\cite{lo2023semanticreaderprojectaugmenting, head2021scholarphi}), and synthesis (e.g.,~\cite{kang_synergi_2023, kang_threddy_2022, wang2024scidasynth}).

However, the accuracy and utility of reviews can decay over time~\cite{Shekelle2001ValidityOT, elliott2017living}, leading to interest in the potential of ``living reviews'' that are continually updated, incorporating relevant evidence as it becomes available~\cite{elliott2014living}. Organizations such as Cochrane\footnote{\url{https://www.cochrane.org/}} offer guidance toward living systematic reviews (LSRs)~\cite{elliott2017living}, contributing a series of works that provide context and justification for LSRs~\cite{elliott2017living}, reason about the role of technology in supporting the creation of LSRs~\cite{thomas2017living}, develop methods for updating the statistical meta-analyses found in LSRs~\cite{simmonds2017living}, and outline the potential for living guidelines that offer dynamic recommendations based on LSRs~\cite{akl2017living}.

Despite this guidance on the conduct, reporting, and publication of LSRs~\cite{cochrane2019handbook, elliott2017living}, studies suggest practical challenges in sustaining these workflows have limited the success of such living reviews~\cite{tricco2008following, heron2023update}. One recent analysis of LSRs on COVID-19 found that most LSRs were never updated after their initial publication, underscoring the overall difficulty of keeping such reviews alive if reliant on largely manual updating efforts~\cite{heron2023update}. Similarly, in climate science, US federal law mandates a process of updating a summary of relevant scientific knowledge every four years. This process is laborious and costly, with the most recent update involving over 500 authors.\footnote{\url{https://nca2023.globalchange.gov/}} While AI tools are being considered as aids for this updating process, they have yet to be adopted~\cite{AlKhourdajie2024TheRO}. Overall, existing guidance on LSRs appears to insufficiently recognize the potential for technological support in what remains a predominantly manual review updating process, and the aging considerations ought to be reexamined in light of recent AI advancements.

Moreover, while existing work has focused on living \textit{systematic} reviews, supporting the process of updating semi-systematic or narrative reviews has received little attention.
Systematic reviews which adhere to explicit search protocols and often integrate findings through meta-analysis require distinct considerations for incorporating new evidence. In contrast, narrative reviews adopt less formal search and appraisal methodologies and serve broader aims of overviewing a research area or informing an agenda for further research. Our study examines the narrative review authoring process, exploring how and when these reviews created with semi-systematic methods should be updated.


%% file: sections/03_methods.tex
\begin{table*}[t]
\centering
\small
\renewcommand{\arraystretch}{1.2} 
\caption{Participant biographies and self-reported research expertise.}
\begin{tabular}{cp{0.2\textwidth}p{0.4\textwidth}}
\toprule
\textbf{ID} & \textbf{Biographical Information} & \textbf{Research Expertise} \\ 
\midrule
P1 & 40, male, assistant professor & Advanced cyber systems, digital equity, distributed systems \\
P2 & 26, male, PhD student & Critical HCI, queer HCI \\
P3 & 29, male, PhD student & Hardware security \\
P4 & 33, female, PhD student & Safety technologies, transformative justice \\
P5 & 27, male, PhD student & Persuasive interventions for absent-minded smartphone use \\
P6 & 28, male, PhD student & Low-power wide-area network, LoRa networking \\
P7 & 27, male, PhD student & Signal processing, video anomaly detection \\
P8 & 35, female, assistant professor & Human-computer interaction, realism, audio, games \\
P9 & 33, female, PhD student & Algorithmic systems in housing services \\
P10 & 36, male, assistant professor & Human aspects of software engineering \\
P11 & 33, male, research engineer & Knowledge distillation, NLP, CV, IR, symbolic regression \\
\bottomrule
\end{tabular}
\label{tab:participant_demographics}
\end{table*}

\section{Methods}
We conducted in-depth, retrospective interviews with authors of narrative reviews, drawing on their expertise to inform the updating process. To identify authors, we searched dblp\footnote{\url{https://www.dblp.org}} for peer-reviewed narrative reviews (equivalently, \textit{survey articles} or \textit{surveys}) in ACM Computing Surveys (CSUR), a premier journal for surveys in computing research. We identified additional surveys published in ACM CHI and CSCW using the search query ``review OR survey.'' We filtered surveys to those published 1-3 years prior, allowing time for new research to emerge while ensuring participants could recall and reflect on their processes. We then excluded any that were not actual surveys of computing literature, for instance papers on improving methodologies for online user surveys or analyzing app reviews. Due to IRB restrictions, we excluded papers whose corresponding authors were based outside the United States or Canada. This process yielded a total of 128 surveys, and we invited each corresponding author to participate via email.


\subsection{Participants}
Eleven of the invited authors participated in our study and were included in the subsequent analysis. Six of their surveys were published in CSUR, four in CHI, and one in CSCW. The surveys were published between early 2022 and 2024, with the majority published in 2023. Due to the lengthy peer review and publication process, most participants noted the relevant research work occurred one to two years prior to official publication. Participants' ages ranged from 26 to 40 ($M = 32, SD = 4.5$). Three were female and eight were male. Three were research tenure-track assistant professors, seven were PhD students, and one was a research engineer. There was considerable diversity among participants, both demographically and in survey article topic (Table~\ref{tab:participant_demographics}). We provide participants' self-identified areas of research expertise (which align closely with the topics of their survey articles they were invited to discuss) but refrain from disclosing the surveys to maintain confidentiality. We anticipated our study size would allow for sufficient theoretical saturation, and that the variety of participants' demographic and research backgrounds would allow us to elicit a rich diversity of survey authoring and updating practices.

\subsection{Interview Protocol}
Our interviews consisted of open-ended questions prompting participants to recall and reflect on their research practices. For instance, participants were asked to describe how they conceived of their survey topic, the composition and responsibilities of their research team, and their approaches for discovering and screening papers, extracting and organizing information, writing, and revising. Probing questions were used to elicit specific facets of participants' experiences and encourage more detailed discussion around particularly meaningful or evocative aspects of their process. Additional details are available in Appendix~\ref{sec:interview_protocol}. Where appropriate, participants were encouraged to share their screen to illustrate their recalled experiences and any artifacts they created. Interviews were conducted virtually via Google Meet, each lasting approximately one hour. Participants were compensated with a \$40 USD gift card. This study protocol was approved by the IRB at the research team's university.

\subsection{Analysis}
Data for analysis included transcripts automatically created from the interview audio recordings and manually cleaned. All personal identifiers were removed from the transcripts to ensure anonymity. We then followed a thematic analysis approach~\cite{braun_thematicAnalysis_2006} to analyze the data. Two team members reviewed all transcripts to familiarize themselves with the data, and then assigned specific codes to two transcripts using an open-coding approach~\cite[Chapter 8]{strauss1998basics}. An initial codebook was formed by merging these codes and refined through discussion. To align divergent observations, coders identified overlapping codes, clarified code definitions, and excluded codes not aligned with the research objectives. Each coder then independently applied the established codes to half of the remaining data. Emergent higher-level themes were discussed and iteratively induced from the codes, informing the following findings.

%% file: sections/04_findings.tex
\section{Findings} \label{sec:findings}
We first summarize the processes and challenges participants described across four core survey authoring activities. We then examine their motivations, strategies, and obstacles for updating their surveys. Finally, we highlight their perspectives on the potential role of AI in aiding survey authoring and updating.

\subsection{The Processes in Authoring Survey Articles: Work Practices and Challenges}

\paragraph{\textbf{Search}}
Participants described a diversity of strategies for paper discovery, some more systematic than others. For example, most mentioned using a scholarly database to find papers, though several participants highlighted the challenge of forming an effective search query. P4 illustrated a lengthy process of iteratively searching for papers with an initial query based on his own expertise, scanning relevant papers returned by the initial query to identify other relevant keywords, and repeating the process until saturation. Another participant found it challenging to identify the right keywords to search for, especially in emergent research areas that lacked a consistent vocabulary (P8). 

Participants also mentioned using more informal methods of paper discovery, including citation chaining~\cite{webster2002analyzing} from foundational papers, receiving recommendations from their social network, and monitoring prominent research groups in their field. Two participants had no dedicated means of paper discovery, relying instead on a collection of papers gathered throughout their research, and searching for and adding new papers only when necessary during the writing process (P1, P7).

\paragraph{\textbf{Appraisal}}
After collecting a set of papers, participants described carefully screening those papers for inclusion. Though most could be filtered based on their title and abstract alone, some participants noted how certain papers required more careful inspection of the introduction, implementation, or results to determine relevance. As with paper discovery, emergent research areas complicated paper screening, as P6 described how the lack of standardized terminology made it challenge to ``comprehend the essence of what they were trying to say and implement.''

Participants also noted a challenge in identifying and removing duplicated papers (P1, P3, P5, P8, P10). Some papers could have multiple versions, e.g., a pre-print, a conference paper, and a journal article, yet offer the same contributions. De-duplication quickly became ``very annoying'' and ``tedious'' over a large corpus (P10). Other strategies appeared more arbitrary and driven by experience. For instance, several participants further filtered ``pseudo-duplicate'' papers; P1 described this aspect of his screening process as more art than science:
\begin{quote}
    \textit{``I found a lot of semi-duplicates. There are some authors who do double dipping. They would write two papers, but they are based essentially on the same prototype. But this is more like art than an algorithm because sometimes I suspected authors use the same prototype, but I could not guarantee it. I tried to remove everything that I would suspect was either not really implemented or was double dipping. I tried to only keep the papers and projects that I knew that actually did something.''}
\end{quote}
However, these subjective processes rely on the authors' expertise and are rarely made explicit in the survey article, making them inherently opaque, open to interpretation, and potentially challenging to replicate when updating the survey in the future.

\paragraph{\textbf{Synthesis}}
After screening papers for inclusion based on relevance and quality, participants often described finding the right organization as the most time-consuming and cognitively challenging phase of their workflow (P1, P3--P6, P10, P11). Organizing hundreds of papers into a multi-dimensional taxonomy was a lengthy process of iterative refinement, sometimes with multiple authors collaborating to reach a consensus on the appropriate structure, dimensions, and evidence. Reflecting on his workflow, P1 described feeling humbled in his taxonomy development process:
\begin{quote}
    \textit{``I had to discard a lot of ideas because initially what I thought it would be just didn't work. It ended up being nonsense. Our initial dimensions, because they were correct but meaningless, they were not informative. It took a lot of time and humility, a lot of humbleness to accept that those initial things didn't work.''}
\end{quote}

Many participants developed a codebook alongside their taxonomy, capturing key questions or dimensions to compare across papers. Predefined codes were based on their surveys' research questions and own expertise, with new codes added as needed. Revisions were labor-intensive, with P10 noting how adding a code meant revisiting every paper and reestablishing a mental model of the paper to extract the relevant information.

\paragraph{\textbf{Interpretation}}
Finally, participants highlighted a key challenge in synthesizing and interpreting the literature. They described how beyond the organizational taxonomy, the real value of a survey lies in its identification of key challenges, future trends, and open research opportunities (P3, P6, P7, P9, P11). P6 emphasized this point, while acknowledging the difficulty in articulating those insights:
\begin{quote}
    ``\textit{Beyond the taxonomy, another important thing about this literature review is the challenges and the future trends you uncovered in the review...because the reason they want to read the paper is because they want to start their own research project. So I did spend a lot of time on these challenges and the trends, on how to make it clear and comprehensive and fancier to give back more insights to those readers.}''
\end{quote}
He elaborated that while the taxonomy he had developed worked well for categorization, it was too rigid for discussing future trends. Providing ``coarse-grained'' insights to inspire new research was ``tricky'' since they were not directly tied to the ``logic'' imposed by the taxonomy. Deriving meaningful insights can also be challenging due to limited perspective, with P6 adding that identifying emerging trends required insight he gained through discussion with other researchers rather than solely reviewing the literature.
\begin{quote}
    ``\textit{At that time, my vision was still quite limited. I can read lots of papers and try to summarize in the tables, even just write the sections to introduce each work, but it's really hard for me to get the sense in five years, what LoRa networking research would be. It's really hard for me to uncover this kind of future trends just by reading papers. I needed to talk to different researchers working on this research field to understand their vision for this topic. It's just something I couldn't do by myself. That's kind of the most challenging part.}''
\end{quote}

Another challenge involved the selection and interpretation of key papers. In contrast to SLRs in clinical medicine, which use rigid inclusion and exclusion criteria to identify relevant studies for meta-analysis and avoid introducing bias, participants in our study described more informal search and appraisal processes, while aiming to be comprehensive in their overall search of the literature. Furthermore, not all relevant research could be synthesized given the length restrictions of a survey article, as one participant explained his nuanced process of selecting papers that balanced recency so ``readers will find it interesting'' and foundational work to avoid ``forgetting the theory'' (P3). Another participant, P7, included additional meta-commentary, drawing connections between contemporaneous papers with similar contributions and critically evaluating when a paper's proposed approach was unsupported by its evidence.

\subsection{The Dynamics of Updating Survey Articles: Motivations, Approaches, and Obstacles}

\paragraph{\textbf{Motivations}}
Participants viewed keeping survey articles up-to-date as a valuable, albeit costly, endeavor. P10 felt keeping his survey current was important, effectively ``showing a picture of how the research is at that point.'' P11 similarly mentioned how updating his survey would be ``helpful for the community,'' as many researchers used it as a starting point, though he expressed concern about its relevance, acknowledging, ``it's already two years old.'' Participants noted how this perception of value also carried an implicit expectation of currency with the latest research. Some recollected how peer reviewers asked for an update to the paper search, since a year had elapsed while under review and new, relevant research may have emerged (P8, P10). After a subsequent rejection, P8 also proactively updated their survey to include new literature, fearing future reviewers would be ``annoyed'' that the survey article was too ``old and outdated.'' For many participants, this expectation manifested as a pressure to publish surveys as quickly as possible and an anxiety surrounding the potential extra work in updating if the paper were not accepted.

Despite believing that surveys should ideally be kept up to date, participants held varying expectations on the desired frequency of updates. For instance, P8, working in a slower-moving field, noticed only a few relevant papers to their survey each year. In contrast, other participants reported encountering many relevant papers to include through passive monitoring of email alerts, conference and journal proceedings, and work citing their survey (P4, P5, P11).

\paragraph{\textbf{Approaches}}
Participants described three main types of updates they envisioned making to their surveys:
\begin{enumerate}[label=\arabic*.]
    \item \textbf{Empirical}, involving the update of quantitative and qualitative evidence presented in text, tables, and figures.
    \item \textbf{Structural}, involving the update of paper structure, organizational taxonomies, and frameworks.
    \item \textbf{Interpretative}, involving the update of synthesis and interpretation of empirical evidence, such as the discussion of limitations, implications, emerging trends, unresolved and resolved challenges, and future directions.
\end{enumerate}

First, participants identified \textit{empirical} updates that aimed to incorporate the latest research findings. Empirical updates involve revising individual quantitative results to reflect the state-of-the-art, as well as any recalculating statistical meta-analyses and aggregated values, such as counts and proportions of papers within each dimension of the taxonomy. These updates may also involve refining the qualitative narrative by replacing less effective or outdated references with more compelling data, a process P10 paralleled to selecting the best quote from an interview study. One participant further characterized making empirical updates as a complex decision-making process, as each new piece of evidence requires consideration of whether ``\textit{you should replace this, you should add this, or you should just mention it in the table without any text}'' (P6).

Second, participants believed empirical updates could eventually warrant \textit{structural} updates, such as adding a new section to a survey to synthesize recently incorporated papers (P2, P5, P6). One participant described how changes in the momentum of a research area could induce a structural update for his survey:
\begin{quote}
    ``\textit{If some field is becoming more prominent and lots of work is going in a particular direction---maybe using SSVEP now more with VR---then we will highlight it in the discussion, make a section for it, and explain the reason why we think it's gaining prominence now, highlighting some of the results.}'' (P5)
\end{quote}
P6 further explained that the decision to create a new section in a survey---as opposed to integrating the evidence into an existing section---should depend on the sufficiency of new evidence, saying ``\textit{if it’s just five papers for this section for this new topic, I don't think it's worth adding a new section. But if there are 20, 30, even 100, it will be really useful to make them into one independent section}.'' Two participants described a similar type of structural update they made while revising their survey, splitting a dimension of their taxonomy into two as the number of included papers grew beyond a reasonable organization (P2, P11). Lastly, participants noted how structural updates could arise from significant changes to the social or technical status quo. For instance, many participants mentioned large language models as a notable paradigm shift they would incorporate into an updated survey (P3, P5, P6, P9, P11), with proposed revisions ranging from adding a section on their usage to restructuring around the pre- and post-LLM eras.

Third, participants described \textit{interpretative} updates as a process of critically re-evaluating a survey article, aligning it with the current state of a research field. Participants saw this update as the key cognitive challenge, involving re-synthesizing the proposed challenges, trends, and opportunities given the updated empirical evidence. Interpretative updates could further be self-reflective, as P5 explained:
\begin{quote}
    ``\textit{It's not just generating new limitations and recommendations, but also considering these recommendations I already made, and seeing how the field is progressing and whether those limitations are being overcome.}''
\end{quote}
Some participants hoped new research would address prior limitations they had raised in an earlier iteration of their survey, resulting in a discussion of both resolved and emerging challenges in the updated survey (P1, P4, P5).

Participants were generally confident that the original organizational structures and taxonomies they had developed for their survey would remain relevant for several years after its initial publication. P1 highlighted the incremental nature of research he saw:
\begin{quote}
    ``\textit{If they happen to introduce a completely new method, then the methods need to be updated, a new workflow must be developed, which I doubt because I have not seen anything really happening. People just squeeze from the previous one, they try to improve the performance of the existing methods.}''
\end{quote}
Participants therefore viewed empirical and interpretative updates as the bulk of maintaining a living survey, with structural updates necessary only after a critical mass of new work.

Given the perceived stability of their surveys' structure, in considering approaches to updating, participants emphasized the intention to reuse much of their original workflows. Maintaining the same processes would not only conserve effort, but also help ensure consistency in an updated survey. Participants mentioned reuse along logistical aspects---such as reforming the same research team---and mechanistic processes, like reapplying the original paper search criteria across the same scholarly databases, filtering papers with the same inclusion criteria, extracting data from new papers using the existing codebook, and rerunning automated scripts for quantitative analysis.
\paragraph{\textbf{Obstacles}}
Participants identified the lack of strong academic incentives as a key structural barrier to keeping surveys continuously updated. Despite being highly cited and valuable to the research community, participants saw the impact of surveys as tied to their initial publication. Without structural support in the form of academic recognition, participants found it difficult to invest the time and effort required to revise an existing publication. P4 explained how even when authoring their original survey, they needed to ``draw a line'' on the timeframe of surveyed literature, as adding new papers was ``moving the goalposts'' and delayed publication.

In addition to the lack of extrinsic incentives, continuous updating was seen as an ``unmanageable'' without additional support. One primary challenge was the need to continuously discover and screen new research papers. The search had to be comprehensive while avoiding duplication, and certain means of paper discovery, e.g., social recommendation, would be absent once the original survey team dissolved. In contrast, participants saw refining an organizational taxonomy as less challenging when updating a survey, as their previous efforts could largely be reused or adapted.

Participants also noted that updating a survey with new evidence could trigger a cascade of revisions (P1, P8, P10). For example, adding papers to the taxonomy would require recalculating statistics, revising text, updating charts and figures, and modifying conclusions. Managing these interconnections to ensure consistency across a long survey article presented additional cognitive challenges. As a potential coping strategy, P1 suggested he would make updates in bulk, which he felt was more efficient than updating one paper at a time. In sum, while participants recognized the community value of living surveys, they struggled to justify the effort due to the lack of personal academic recognition.

\subsection{Perspectives of AI Use for Survey Updating}

\subsubsection{AI Lacks Nuanced Understanding for Expert-Level Reviews}
Participants expressed concerns about AI’s ability to produce surveys that are as insightful and meaningful as those written by human experts (P1, P2, P5--P8). They emphasized that scholarly synthesis was more than just processing and summarizing data, demanding nuanced understanding, critical engagement, and the foresight to identify gaps and future trends that may advance the field. Several participants noted these aspects to be the core value of survey articles which drive meaningful discourse.

Participants also argued that without domain expertise, AI syntheses are likely to lack the depth and complexity needed for such discourse (P5--P8). P8 highlighted the importance of human oversight to avoid shallow analysis:
\begin{quote}
    ``\textit{I think you can use it. I just think you need to be way more careful about having a human-in-the-loop, and having a human being the final position of authority on what happens during the final part of the analysis. Because otherwise your analysis is going to be incredibly shallow.}''
\end{quote}
This need for experts as a critical voice was echoed by P5, who emphasized that relying too heavily on AI for updates risks making survey articles formulaic. P5 expressed his belief that if articles were written to be updated by AI, they would lack creativity and reader engagement:
\begin{quote}
    ``\textit{If we try to set up the paper in such a way that an AI can easily update sections of it to match data, it may not be an interesting paper to read... And I think papers are written to be fun.}''
\end{quote}
Overall, participants perceived current AI systems as lacking the ability to offer the depth, creativity, and engagement characteristic and required of expert-level survey articles. Without these elements, AI-generated syntheses risked becoming too shallow or formulaic.

\subsubsection{Roles of AI in Supporting Updates}
While participants were generally skeptical about AI's ability to replicate human-written survey articles, they recognized the potential for delegating repetition or tasks requiring less specialized expertise to the AI. Participants recommended using AI for tasks where mistakes would be less costly (P1, P7, P10). For instance, P10 described how he would only choose to use AI in places where ``the results of the paper and the conclusion do not rely very strongly on it.'' P7 suggested that AI should only be tasked with the ``very easy work,'' and for the creative work, such as discussing trends and challenges, AI should only be used to offer ``suggestions.'' The following summarizes three roles participants highlighted of AI assistance throughout authoring and updating a survey.

\paragraph{\textbf{AI for routine automation}} 
First, participants identified routine numerical updates, such as tallying results or updating statistical information, as tasks where AI could be reliably employed (P4, P5). Automating these tasks could help ensure consistency and comprehensiveness when incorporating updates into a survey. For example, participants expressed the need to extract specific data from a corpus of papers to facilitate coding, and then subsequently ensure the quantitative aspects of the survey were updated to reflect the latest codes. One concern participants shared was the need to verify the work done by any AI system (P5, P10, P11). As P5 explained, ``you may not be able to trust the extracted data,'' highlighting the importance of proper guardrails to mitigate overreliance and providing affordances for user verification.

\paragraph{\textbf{AI as a surrogate}}
Participants described specific sub-tasks where AI could be valuable in managing a large corpus of data (P2, P5, P7--P11). For instance, they saw AI as a useful assistant in potentially making the search process more comprehensive by refining and augmenting search strings (P6, P10, P11). For large paper collections, AI was seen as helpful for doing an initial pass at screening and organizing of the papers (P2, P5, P7--P9). However, participants also expressed concerns about AI missing important information or adding irrelevant information. In identifying or updating a corpus of papers, P9 described how it is easier to spot irrelevance but more challenging to detect what may have been overlooked. Furthermore, some participants emphasized the importance of establishing proper ways to document AI usage for reproducibility and transparency in research (P8, P10).

\paragraph{\textbf{AI as a second opinion}}
Participants discussed several places in the authoring and updates process where AI could serve as a co-author providing a second opinion (P1, P5, P6, P9--P11). For instance, when analyzing trends in a cluster of research papers, AI could do a ``sanity check'' in case the researcher misses anything (P9). Similarly, when working alone and screening papers, AI could ``provide an opinion'' alerting the researcher to papers that might require taking a closer look (P5). These examples highlight how integrating AI as a collaborative partner could be beneficial within authoring or updating, to enhance systematicity, reduce cognitive bias, and encourage a more comprehensive and insightful synthesis.

%% file: sections/06_discussion.tex
\section{Discussion}

\begin{table*}[t]
\centering
\small
\renewcommand{\arraystretch}{1.3}
\caption{Opportunities for supporting various updates to narrative literature reviews.}
\begin{tabular}{p{0.1\textwidth}p{0.2\textwidth}p{0.6\textwidth}}
\toprule
\textbf{Update Type} & \textbf{Description} & \textbf{Opportunities for AI Assistance in Updating} \\
\midrule
Empirical & Revision of quantitative and qualitative evidence presented in text, tables, and figures & 
\begin{minipage}[t]{\linewidth}
    \begin{itemize}[leftmargin=*]
        \item Identify and recalculate numerical values or meta-analyses
        \item Update references to evidence in figures or tables
        \item Change or revise representative examples in text
    \end{itemize}
    \vspace{0.3em}
\end{minipage} \\
Structural & Revision of paper organization, taxonomies, and frameworks & 
\begin{minipage}[t]{\linewidth}
    \begin{itemize}[leftmargin=*]
        \item Reevaluate included evidence and definition of an existing taxonomy dimension
        \item Group emerging research to align with or challenge existing taxonomy dimensions 
        \item Identify opportunities to improve paper organization (e.g., by splitting sections that exceed a threshold of research)
    \end{itemize}
    \vspace{0.3em}
\end{minipage} \\
Interpretative & Revision of evidence synthesis, interpretation, and narrative framing & 
\begin{minipage}[t]{\linewidth}
    \begin{itemize}[leftmargin=*]
        \item Identify potential biases, assumptions, and narratives in the original survey that may be validated, challenged, or updated in light of new evidence or technologies
        \item Analyze the extent to which prior gaps and limitations in the research have been addressed
        \item Propose new unresolved challenges or under-explored areas
    \end{itemize}
    \vspace{0.3em}
\end{minipage} \\
\bottomrule
\end{tabular}
\label{tab:updates_and_ai_support}
\end{table*}

\subsection{Summary of Findings}
Our study examined the work practices within survey article authoring and the challenges researchers foresee in maintaining their currency. Our findings reiterate four key processes of review---search, appraisal, synthesis, and interpretation---and highlight the many inherent subjective, expertise-driven decisions and idiosyncratic strategies. Participants noted the value in keeping their surveys updated for the research community and described three main types of updates they would make: empirical, structural, and interpretative. Participants characterized updating as a process of restoring institutional knowledge and attempting to adapt and reuse their original workflows. For instance, new research can be identified through established search methods and integrated into an existing taxonomy, while sufficient research could further motivate deeper structural and interpretative revisions.

Despite the perceived value, participants also expressed how continuous survey updating was infeasible and misaligned with existing academic incentives. Many therefore recognized the potential for AI support to lower the costs of updating, such as by helping to identify new, relevant papers or automating repetitive information extraction tasks, though skeptical of its ability to replace their own learned expertise for synthesizing evidence. Our findings, while focused on scholars in the computing field, align with perspectives from other empirical studies involving scholars across diverse disciplines~\cite{messeri2024artificial, morris_scientists_2023, chubb2022speeding}. For instance, \citet{messeri2024artificial} find that objective tasks are more conducive to helping scholars establish appropriate trust in AI assistance, and several other studies similarly suggest scholars are more receptive to AI's role in narrow tasks that boost personal productivity but are hesitant to rely on AI in ``emotional tasks'' that require creativity and complex decision-making. For these nuanced tasks, participants instead see AI as suitable to ``augment and assist human judgment''~\cite{chubb2022speeding}.

\subsection{Toward a Vision of Living Narrative Reviews}

\subsubsection{Considerations for AI-Assisted Updating of Narrative Reviews}
Our findings offer several considerations for how future AI systems could aid updating workflows. First, systems could help determine \textit{when} an update is needed by continuously monitoring new research and identifying those relevant to existing taxonomies and syntheses. They could further explain \textit{why} new research is relevant, such as by highlighting how it aligns or challenges existing parts of a survey. Updating efficiency could be improved by localizing \textit{where} updates are needed and directing authors' attention to specific components---text, figures, or tables---that should be revised. This could be especially helpful for cascading revisions, as participants noted, since integrating even a single paper may require updating many parts of a survey, such as where it is introduced, synthesized with related research, and visualized within tables or figures. Finally, systems could offer guidance on \textit{how} to integrate new research into a survey, for instance by identifying the type of update required---empirical, structural, or interpretive---and providing tailored revision recommendations.

These three different types of updates in narrative reviews also present unique challenges that AI systems could help address. For empirical updates, systems could help recalculate numerical values and meta-analyses given new research, where providing transparency to allow researchers to verify the updated evidence is important. More challenging is supporting potential structural updates, such as determining when new research is sufficient to extend or revise a taxonomy and its dimensions, which participants felt required a deep and nuanced understanding of how new research may reshape the overarching conceptual framing of a research area. Future AI systems could also likely aid in interpretative updates, for instance by meaningfully synthesizing literature within a taxonomy dimension, reevaluating gaps in a research area, or informing an evolving agenda for future research. These technologies are promising yet nascent, with recent studies echoing participants' concerns regarding the shallow nature of AI-generated literature syntheses~\cite{martinboyle2024shallow}. Overall, narrative review updating remains for now a collaboration of AI and human effort, though given the high costs and low incentives of making frequent updates, realizing living narrative reviews may eventually require a concentration of human effort on verifying and steering AI-assisted updates.

\subsubsection{Systematicity and Expertise in Narrative Reviews}
Designing future systems to enable living narrative reviews requires balancing systematicity with the subjective, expertise-driven nature of survey authoring. While evidence-based disciplines like clinical medicine emphasize systematic methodologies---such as transparent, repeatable methods for identifying, appraising, and synthesizing literature---these standards may not fully align with semi-systematic methods such as in computing research surveys. Participants described workflows marked by subjective decisions informed by years of specific research expertise, across literature discovery, taxonomy creation, synthesis, and narrative construction. These choices, shaped by implicit knowledge and anticipation of reader needs, were rarely documented in surveys themselves, posing challenges for both subsequent human-driven and AI-supported updates. Without sufficient systematicity or documentation, restoring the institutional knowledge of a research team or replicating the original idiosyncratic processes to perform consistent updates becomes significantly harder.

Future AI systems may bridge this gap by supporting authors in externalizing the implicit strategies underlying their subjective decisions. Interactive tools could invite authors to articulate their search and synthesis methodologies more explicitly, clarify decision-making criteria, and document expertise-driven strategies, such as when participants described targeting specific journals or research groups. An exciting direction for future research lies in the design of mixed-initiative systems to then leverage these explicit methodologies and infer patterns in existing surveys to provide updating assistance aligned with the author’s narrative and methodological intent, while making transparent any inferred subjective choices. Altogether, these approaches can serve to encourage systematicity in survey authoring and ensure subsequent updates to a survey are efficient, reliable, and consistent with prior survey iterations.

\subsection{Limitations and Future Work}
The retrospective nature of the interviews could introduce recall bias, as participants may have selectively reported the most salient challenges, overlooking others. Future work could use more direct observational methods, such as contextual inquiry, to capture authors’ real-time processes as they update their literature reviews. To complement authors' perspectives on continuously updated reviews, additional studies could investigate \textit{readers}’ needs. For instance, what form should a living review take and how might new affordances and interactions facilitate their use? Similar techniques could also aid peer review of review updates, which remains an open challenge for realizing living reviews.

%% file: sections/07_conclusion.tex
\section{Conclusion}
We conducted retrospective interviews with 11 authors of computing survey articles to examine their practices for creating and updating narrative reviews. We identified three main types of updates---empirical, structural, and interpretative---each with unique motivations, challenges, and opportunities for AI assistance. Although participants supported the idea of continuously updated reviews, they were skeptical about its feasibility due to current academic incentives, suggesting the important role AI tools may serve in facilitating or automating these updates. This study takes a first step toward enriching our understanding of when, why, and how these reviews should be kept up-to-date, informing the development of future tools to enable living narrative reviews.

%% file: sections/08_appendix.tex
\section{Interview Protocol} \label{sec:interview_protocol}
Our semi-structured interviews included the following guide questions. When appropriate, follow-up questions were used to encourage participants to elaborate on their responses, for instance to probe deeper into process details or recall motivations for a particular decision. In \textit{Process.}, participants were asked to describe their workflows in as much detail as possible. Guide questions for specific scholarly activities were used only if unacknowledged by participants; otherwise, the discussion proceeded naturally, e.g., ``\textit{What did you do next?}''

\begin{itemize}
    \item \textit{Introduction.} Could you briefly describe your area of research and the survey paper?
    \item \textit{Team Composition.} Could you describe the research team involved? How large was the team? Broadly, what were the responsibilities of each member?
    \item \textit{Process.} Could you walk me through your entire process of writing this survey paper, starting from the beginning?
    \begin{itemize}
        \item \textit{Ideation.} How did you come up with the idea for the survey paper?
        \item \textit{Paper Search.} Could you walk me through your initial paper search process? For instance, how did you find, filter, or organize papers?
        \item \textit{Coding.} How did you begin to make sense of the information within the papers?
        \item \textit{Writing.} How did you translate the information you collected and organized into writing the paper? 
    \end{itemize}
    \item \textit{Perceived Challenges.} What did you feel were the most challenging aspects of the process?
    \item \textit{Tedious Aspects.} Were there any parts of the process that stood out as tedious, time-consuming, or cognitively challenging?
    \item \textit{Intermediate Artifacts.} What intermediate products (e.g., documents, spreadsheets) did you create, if any, during the process? How did you use those artifacts? What was helpful or not helpful about these artifacts?
    \item \textit{Tools.} What tools did you use throughout the process of writing this paper? What worked well, and what could have worked better?
    \item \textit{Process Adjustments.} If you were to write this survey paper again, what would you have done differently?
    \item \textit{Update Process.} Could you walk me through how you would update this paper? (Including similar sub-questions as in \textit{Process.} above.)
    \item \textit{Update Elements.} Which parts of the paper do you envision would change from these updates? For instance, could you speak to potential revisions regarding the existing text, structure, figures and tables, or other elements?
    \item \textit{Update Challenges.} What do you believe are the main challenges in the process of updating this paper?
    \item \textit{Desired Support.} What kind of support would you want in the process of updating this paper?
    \item \textit{AI Perspectives.} Which parts of the survey paper authoring or updating process do you believe an AI would be capable or incapable of assisting you with? 
\end{itemize}